\documentclass[10pt]{article}

\usepackage{amsmath}
\usepackage{amssymb}
\usepackage{graphics}

\title{\textbf{Towards Quantum Gravity: A Framework for Probabilistic Theories with Non-Fixed Causal Structure}}

\author{Lucien Hardy\\
\textit{Perimeter Institute,}\\
\textit{31 Caroline Street North,}\\
\textit{Waterloo, Ontario N2L 2Y5, Canada}}

\begin{document}

\maketitle

\begin{abstract}
General relativity is a deterministic theory with non-fixed causal structure.  Quantum theory is a probabilistic
theory with fixed causal structure.  In this paper we build a framework for probabilistic theories with non-fixed
causal structure.   This combines the radical elements of general relativity and quantum theory.   We adopt an
operational methodology for the purposes of theory construction (though without committing to operationalism as a
fundamental philosophy).  The key idea in the construction is {\it physical compression}.  A physical theory
relates quantities.  Thus, if we specify a sufficiently large set of quantities (this is the compressed set), we
can calculate all the others.  We apply three levels of physical compression. First, we apply it locally to
quantities (actually probabilities) that might be measured in a particular region of spacetime. Then we consider
composite regions. We find that there is a second level of physical compression for the composite region over and
above the first level physical compression for the component regions.   Each application of first and second level
physical compression is quantified by a matrix. We find that these matrices themselves are related by the physical
theory and can therefore be subject to compression. This is the third level of physical compression. This third
level of physical compression gives rise to a new mathematical object which we call the causaloid. From the
causaloid for a particular physical theory we can calculate everything the physical theory can calculate. This
approach allows us to set up a framework for calculating probabilistic correlations in data without imposing a
fixed causal structure (such as a background time).  We show how to put quantum theory in this framework (thus
providing a new formulation of this theory). We indicate how general relativity might be put into this framework
and how the framework might be used to construct a theory of quantum gravity.
\end{abstract}

\section{Preliminary remarks}

The great outstanding problem in theoretical physics left over from the last century is to find a theory of quantum
gravity. A theory of quantum gravity (QG) is a theory which approximates quantum theory (QT) and general relativity
(GR) in appropriate limits (including, at least, situations where those theories have already been experimentally
verified).  The problem is to go from two theories which are less fundamental to one which is more fundamental.  Of
course, it is possible, at least logically, that a theory of quantum gravity can be entirely formulated inside one
of these two component theories.  The main approaches to QG assume that the quantum framework is sufficient.
Indeed, it is often stated that the problem is to quantize general relativity.   In string theory (and its various
descendants) an action is written down which defines the motion of strings (or membranes) on a fixed spacetime
background \cite{strings}. This is formulated entirely within the quantum framework.  In loop quantum gravity
Einstein's field equations are written in cannonical form (so we have a state across space evolving with respect to
some time parameter) and then quantization methods are applied \cite{loops}.  In this paper we will not assume that
QG can be formulated entirely within the standard quantum framework.  Rather we will take a more evenhanded
approach. We note that both GR and QT have conservative and radical features.
\begin{description}
\item[General Relativity]
\begin{description}
\item[]
\item[Conservative feature:]  General relativity is deterministic.  Given sufficient information on a boundary,
there is a unique solution for the physical observables in the theory.
\item[Radical feature:]  The causal structure is non-fixed.  Whether a particular interval $\delta x^\mu$ is
spacelike or timelike is not specified in advance but can only be determined once we have solved the Einstein field
equations for the metric.
\end{description}
\item[Quantum Theory]
\begin{description}
\item[]
\item[Conservative feature:] The causal structure is fixed in advance.  We will elaborate on this in Sec.\
\ref{explorationof} below.
\item[Radical feature:]  The theory is irreducibly probabilistic.  That is to say, we cannot state the postulates
of standard QT without reference to probabilities.
\end{description}
\end{description}
It is curious that each theory is radical where the other is conservative.  It is likely that QG must be radical in
both cases.  Thus, we take as our task to find a framework for physical theories which
\begin{enumerate}
\item Is probabilistic.
\item Admits non-fixed causal structure.
\end{enumerate}
If we are able to find such a framework then we can hope to formulate both QT and GR as special cases.  And, more
importantly, we can expect that QG will also live in this framework.

To begin we need a starting point.  Fortunately, if we look back to the historical conceptual foundations of GR and
modern QT we see that they have in common a certain operationalism.  In his 1916 review paper \lq\lq The foundation
of the general theory of relativity" Einstein motivates the crucial requirement of general covariance in various
ways by appealing to operational reasoning.  For example, he says
\begin{quote}
All our space-time verifications invariably amount to a determination of space-time coincidences. (\dots) Moreover,
the results of our measurings are nothing but verifications of such meetings of the material points of our
measuring instruments with other material points, coincidences between the hands of a clock and points on the clock
dial, and observed point-events happening at the same place and the same time .

The introduction of a system of reference serves no other purpose than to facilitate the description of the
totality of such coincidences \cite{Einstein}.
\end{quote}
(and hence, since these coordinates are merely abstract labels, the laws of physics must be invariant under general
coordinate transformations). The first sentence of Heisenberg's 1925 paper \lq\lq Quantum-theoretical
re-interpretation of kinematic and mechanical relations", which marked the birth of modern quantum theory, reads
\begin{quote}
The present paper seeks to establish a basis for theoretical quantum mechanics founded exclusively upon
relationships between quantities which in principle are observable \cite{Heisenberg}.
\end{quote}
Heisenberg was, of course, very much influenced by the operationalism of Einstein. Given this common starting point
for the two theories, it makes sense to adopt it here also. Thus, we will adopt an operational methodology. Before
proceeding, it is important to qualify this. We are adopting an operational methodology {\it for the purposes of
theory construction}.  This does not commit us to operationalism as a fundamental philosophy (in which the reality
of anything beyond the operational realm is denied). Operationalism is a potentially powerful methodology precisely
because it can remain neutral about what is happening beyond the operational realm and consequently enable us to
make statements about a physical situation we know, at least, are not wrong.

We will try to be particularly careful to formulate a version of operationalism that is useful for our purposes.
The key aspect of the operational realm is that it is possible to accumulate data recording the settings of the
instruments and the outcomes of measurements.   Hence, our starting point will be the following assertion
\begin{description}
\item[Assertion:] A physical theory, whatever else it does, must correlate recorded data.
\end{description}
Of course, a physical theory may do much more than correlate data - it may provide an explanation of what happens,
it may provide a picture of reality, it may provide a unified description of diverse physical situations. However,
in order that a physical theory be considered as such, it must be capable of correlating data. Once again, it is
important to assert that this does not commit us to an operational philosophy of physics. Nevertheless, the fact
that physical theories must be capable of correlating data places constraints on the mathematical structures that
can serve as such theories.   We will look at how a theory can correlate data and find a general mathematical
framework for physical theories.   Operationalism can be regarded as a kind of conceptual scaffolding used to
construct this mathematical framework.  Once we have found this framework we are free, should we wish, to disregard
the scaffolding and regard the mathematical framework as a fundamental description of the world. Something like
this happened when we went from Einstein's operationally formulated version of special relativity to Minkowski's
picture.

In both GR and QT there is a matter of fact as to whether a particular interval is timelike or not (in GR we can
only establish this after solving for the metric). In QG we expect the causal structure to be non-fixed as in GR.
However, in standard quantum theory, there is no matter of fact as to the value of any non-fixed physical quantity
unless it is measured (or specially prepared).  Hence, in QG we expect that there will be no matter of fact as to
whether a particular interval is timelike or not unless a measurement is performed to determine it.  This means
that we cannot assume that there is some slicing of spacetime into a time ordered sequence of spacelike
hypersurfaces. Many of the concepts we usually take for granted in physics, such as evolution, state at a given
time, prediction, and preparation have to be re-examined in the light of these considerations.

The formalism presented in this paper first appeared in \cite{Hardy1}.  The present paper is almost self contained
though a few proofs which do not appear here are in \cite{Hardy1}.

\section{Exploration of causal structure in QT}\label{explorationof}

In this section we will elaborate, as promised, on the nature of the fixed causal structure in QT.  The most
immediate manifestation of this is that the state in quantum theory is given by
\begin{equation}
|\psi(t)\rangle = U(t) |\psi(0)\rangle
\end{equation}
We see that there is a background time $t$ which assumes that there is a certain fixed causal structure (past
influences future) acting in the background.   However, a deeper insight into causal structure in QT is gained by
thinking about the relationships between operators that pertain to distinct spacetime regions. If these two
spacetime regions are spacelike separated then the operators should commute.   In this picture we are thinking of
operators which act on the global Hilbert space. An alternative way of thinking is to imagine a local Hilbert space
corresponding to each spacetime region.   To be more specific consider two spatially separated quantum systems with
Hilbert spaces ${\cal H}_1$ and ${\cal H}_2$ of dimension $N_1$ and $N_2$ respectively. Let system 1 be acted upon
by a quantum gate $A$. Let system 2 be acted upon sequentially by three gates $B$, $C$, and $D$ where gate $B$ is
spacelike separated from gate $A$. Denote the quantum operators associated with the evolution due to each gate by $
A$, $B$, $C$, and $D$ (these operators pertain to the local Hilbert space of the corresponding system). Gates $A$
and $B$ are spacelike separated. Hence the appropriate way to combine the operators $ A$ and $ B$ is to use the
tensor product giving ${A}\otimes{B}$. (As an aside note that the property that the global operators should commute
follows if we write ${a} = {A}\otimes {I}$ and ${b} = {I}\otimes {B}$ for the global operators where $ I$ is the
identity.) Gates $B$ and $C$ are timelike separated and immediately sequential. Therefore the appropriate way to
combine the operators ${B}$ and ${C}$ is by the direct product (composition) $CB$. Gates $B$ and $D$ are timelike
separated but not immediately sequential. The right way to combine operators $ B$ and $ D$ is to use what we will
call the question mark product $[{D}?{B}]$. The question mark product is defined by $[{D}?{B}]{C}\equiv {D}{C}{B}$.
It is clearly a linear operator.  We see that we have here three different products.  To choose the correct one we
need to know, in advance, what the causal relation is between the two regions.  We can only do this if we specify a
particular causal structure in advance and hence this causal structure must be fixed.  We will find a new product -
the causaloid product - which unifies these three types of product treating them in the same way in the context of
a more general framework.  This will enables us to formulate a framework in which, in general, we do not need to
specify in advance whether a particular separation is timelike or spacelike (and, indeed, there may be no matter of
fact as to whether the separation is timelike or spacelike).

To gain a clue as to where this framework will come from consider the above example further.  If we are given
$A\otimes B$ then we can deduce $A$ and $B$ separately.  Likewise if we are given $[D?B]$ we can deduce $B$ and $D$
separately.  This second case is not so obvious - physically what is happening is that it is possible to break any
tight correlation between $B$ and $D$ by considering different possible $C$'s.  In these two cases, all the
information available in the operators before they are combined remains available afterwards. However, if we are
given the operator $CB$ we cannot deduce $C$ and $B$ separately.  The best way to understand the reason for this is
that we can deduce the state for region $CB$ from measurements on region $B$ alone (since it is the same qubit
which passes, in sequence, through these two regions). Consequently, there is a reduction in the number of
parameters required to specify the state for this composite region (unlike in the case of region $AB$). This is
reflected by a reduction in the number of parameters required to represent operators $CB$ in the dual space.  The
reduction in the number of parameters required to specify the operator is due to correlations between the two
regions coming from the physical theory itself (quantum theory in this case).  There is a certain kind of {\it
physical compression}. This is also the only case when there is a direct causal connection between the two regions
- within the context of this quantum circuit there is nothing that can be done to break the correlation between the
two regions. We will say that the two regions are {\it causally adjacent}. Hence we see that {\it causal adjacency
is associated with a certain kind of physical compression} (which we will call {\it second level} physical
compression). It turns out that physical compression is the key - it is the mathematical signature of causal
structure.  Physical compression arises since the physical theory relates quantities and consequently we can have
full information about all quantities by listing a subset (the remaining quantities being deduced from this subset
using relations deduced from the physical theory). We will use the notion of physical compression to formulate a
general framework for probabilistic theories which do not require fixed causal structure.

\section{Collection and analysis of data}\label{collectionand}

In experiments we collect data.  Data consists of (i) a record of actions taken (such as knob settings) and (ii)
results of measurements and observations (for example observing that a detector clicks, or observing the reading of
a clock).  Typically, we will take note when data that is recorded in close proximity.   For example, we might note
that, at time 02:52 according to a clock $A$ which is proximate to the Stern-Gerlach apparatus $B$ which was set at
angle $55^\circ$, the detector corresponding to spin up clicked.  Here we have three pieces of data (02:52,
$55^\circ$, and spin up) all recorded in proximity. We will assume that such proximate data is recorded on a card
(one card for each set of proximate data). Thus, at the end of an experiment, a stack of cards will be accumulated
where, on each card, proximate pieces of data are written as in this example.  Of course, it is not necessary that
cards are actually used - the data could be stored in a computers memory, in a lab book, or in the brain of the
experimentalist. However, the story with the cards will help us set up the framework we are after. The notion of
proximity is clearly a slightly vague one.  On this matter, Einstein \cite{Einstein} writes \lq\lq We assume the
possibility of verifying ... for immediate proximity or coincidence in space-time without giving a definition of
this fundamental concept."  It will ultimately boil down to a matter of convention and judgement as to what data
counts as proximate.  The convention aspect is under our control.  Typically no two events will be exactly
coincident so we will have to set a scale. If the two events occur to within this scale then we will say that they
are proximate. The choice of scale is a convention. So long as we stick with a consistent convention then there is
no problem. However, there is still a matter of experimental judgement in asking whether two events are proximate
to within this scale. The judgement aspect is not so much under our control.

We will assume that the first piece of data, $x$, on each card corresponds to some observation which we will regard
as specifying location. We will have in mind that this corresponds to space-time location (although it is not
strictly necessary that this is the case). For example, $x$ could be the space-time location read off some actual
physical space-time reference frame. It could be the GPS position given by the retarded times of four clocks
situated on four appropriately moving satellites. The remaining data on a card is a record of actions (e.g. knob
settings) and observations.   The choice of actions is allowed to depend on $x$.  A simple example is where the
data stored on the card is of the form $(x, F(x), s)$ where $x$ is the location data, $F(x)$ represents the choice
of actions such as knob settings (this depends on $x$) and $s$ represents outcomes of measurements (for example
spin measurements).  In the case that there are multiple knobs $F$ is a multivariable object, and if there are
multiple measurement outcomes obtained at this location then $s$ is, likewise, a multivariable object. We could
consider more complicated examples such as $(x, r, F(x, r), s)$ where $r$ is data that is not regarded as part of
the data representing location but on which the choice of action can, nevertheless, depend. We will illustrate
these ideas with two examples
\begin{description}
\item[Probes drifting in space.]  We imagine a number of probes ($n=1, 2,\dots$) drifting in space.  Each
probe has a clock with reading $t_n$, some knobs which control the settings, $F(n, x)$, of various measurements
(such as Stern Gerlach orientations) and some meters with readings $s_n$.  At each tick of the clock on each probe
we record on a separate card
\[
(x\equiv(t_n, \{t_m^n\}), n,  F(n, x), s_n)
\]
where $\{t_m^n\}$ represents the retarded times seen at probe $n$ on the other probes (we could choose just a
subset of the other probes here). At the end of the experiment, we will end up with one card for each tick of each
probe.
\item[Sequence of spin measurements.]  Imagine a sequence of five spin measurements performed on a single spin half
particle emitted from a source.  We label the spin apparatuses $x=1$ to $x=5$ and the source $x=0$.  At the source
we collect a card with data $x=0$ followed by whatever data is recorded corresponding to the proper functioning of
the source.  At each spin measurement we collect the data $(x, \theta(x), s)$ where $\theta$ is the orientation of
the spin measurement and $s$ is the outcome (spin up or spin down).  At the end of the experiment we will have a
stack of six cards.
\end{description}
There are many different possible choices for the function $F$ (corresponding to the various possible choices of
knob settings at different locations). We will imagine that the experiment is repeated for each possible function.
Further, since we are interested in constructing a probabilistic theory, we will assume that the experiment can be
repeated many times for each $F$ so that we can construct relative frequencies.  We will imagine that each time the
experiment is performed the cards are bundled into a stack and tagged with a description of $F$.  After having
repeated the experiment many times for each $F$ we will have a large collection of tagged bundled stacks of cards.
To usefully repeat the experiment it may be necessary to reset some aspects of the setup such as the clocks.  The
notion of repeating the experiment is problematic assumption if we are in a cosmological setting.  An alternative
approach is discussed in \cite{Hardy1}.

We will imagine that this collection of tagged bundled stacks of cards is sent to a man inside a sealed room for
analysis. Our task is to invent a method by which the man in the sealed room can analyse the cards thereby
developing a theory for correlating data. The order in which the cards are bundled into any particular stack does
not, in itself, represent recorded data (all recorded data is written on the cards themselves).  Consequently, the
man in the sealed room should not take this into account in his analysis.  To be sure of this we can imagine that
the cards in each stack are shuffled before being bundled.  The order of the stacks also does not represent data
and so we can also imagine that the bundles themselves are also shuffled before being sent into the sealed room.

The usefulness of this story with a man inside a sealed room is that he cannot look outside the room for extra
clues on how to analyse the data. Hence, he will necessarily be proceeding in accordance with an operational
methodology as we discussed earlier. He will have to define all his concepts in terms of the cards themselves. We
will now define some concepts in terms of the cards.
\begin{description}
\item[The full pack,] denoted by $V$, is the set of all logically possible cards over all $x$, all possible
settings and all possible outcomes - any card that can be collected in the experiment must belong to $V$.
\item[An elementary region,] denoted by $R_x$, is the set of all cards taken from $V$ which have some particular $x$
written on them.
\item[A stack,] denoted by $Y$, is the set of cards collected one repetition of the experiment.
\item[A procedure,] denoted by $F$, is the set all cards taken from $V$ which are consistent with the given function
$F$ for the settings.  We intentionally use the same notation for the set and for the function since it will be
clear from the context which meaning is implied and, in any case, the information conveyed is the same (the set $F$
is a more cumbersome way of conveying this information but it will turn out to be useful below).
\end{description}
It is worth noting that we must have $Y\subseteq F\subseteq V$ for a stack $Y$ tagged with procedure $F$. We can
define some more concepts in terms of these basic concepts.
\begin{description}
\item[A region] denoted by $R_{{\cal O}_1}$ is equal to the union of all the elementary regions $R_x$ for which
$x\in {\cal O}_1$.   That is
\begin{equation}
R_{{\cal O}_1} \equiv \bigcup_{x\in{\cal O}_1} R_x
\end{equation}
We will often abbreviate $R_{{\cal O}_1}$ by $R_1$.
\item[The procedure in region] $R_1$ is given by the set
\begin{equation}
F_{R_1} \equiv F\cap R_1
\end{equation}
We will sometimes write this as $F_1$. It conveys the choice of measurement settings in region $R_1$ (more
accurately, it conveys the intended choice of measurement settings).
\item[The outcome set in region] $R_1$ is given by
\begin{equation}
Y_{R_1} \equiv Y\cap R_1
\end{equation}
We will sometimes write this as $Y_1$. It represents the outcomes seen in this region.
\end{description}
Note that
\begin{equation}
Y_1 \subseteq F_1 \subseteq R_1
\end{equation}
These definitions may appear a little abstract.  However, the idea is very simple.  We regard the cards as
belonging to regions, for example $R_1$.   In this region we have
\begin{equation}
(Y_1, F_1) \Longleftrightarrow (\text{outcomes in}~R_1, ~\text{settings in}~R_1)
\end{equation}
We will label each possible $(Y_1, F_1)$ in $R_1$ with $\alpha_1=1, 2, \dots$.  By analysing the cards in terms of
which regions they belong to the man in the sealed room can form a picture of what happened during the experiment.

We are seeking to find a probabilistic theory which correlates data.  It is worth thinking carefully about what
this means.  Probabilities must be conditional.  Thus, we can talk about the probability of $A$ given that
condition $B$ is satisfied.  But even further, the conditioning must be sufficient for the probability to be well
defined.  For example, we can calculate the probability of a photon being detected in the horizontal output of a
polarising beamsplitter given that, just prior to impinging on this beamsplitter, it passed through a polariser
orientated at $45^\circ$ to the horizontal. This probability is well defined (and equal to $\frac{1}{2}$). However,
the probability that the photon will be detected in the horizontal output of a polarising beamsplitter given that,
just prior to impinging on this beamsplitter, it passed through a plane sheet of glass is not well defined.  We
would need more information to be able to calculate this probability.   The lesson to be drawn from this is that it
is not always possible to calculate probabilities.   Thus, we will take as the task of the theory the following
\begin{enumerate}
\item To be able to say whether a probability is well defined.
\item If the probability is well defined to be able to calculate it.
\end{enumerate}
The first task is important and deserves further discussion.  One way to think of this is to adopt an adversary
model.  Thus, imagine that we were to write down a certain probability for a photon being detected in the
horizontal output of a polarising beamsplitter given that it had just passed through a plane sheet of glass.
Whatever probability we write down, we could imagine some adversary who can ensure that this probability is wrong.
For example, before the photon impinges on the sheet of glass, the adversary may send the photon through a
polariser set at some angle he chooses such that our probability is wrong.   However, when we have sufficient
conditioning an adversary cannot do this.  This is clear in the first example where the photon passes through a
polariser set at $45^\circ$ just prior to impinging on the polarising beamsplitter.   How do we usually know
whether a probability is well defined in physical theories?  A little reflection will reveal that {\it we usually
know this by reference to some underlying definite causal structure}.  For example, if we have a full specification
of the boundary conditions in the past light cone of some region and we know what settings are chosen subsequent to
these boundary conditions in this past light cone , then we can make well defined predictions for the probabilities
in that region. However, in the case that we do not have some well defined causal structure to refer to, we cannot
proceed in this way.  In the causaloid framework to be presented we will provide a more general way to answer the
question of whether a probability is well defined.

In the notation above, we wish first to know whether the probability
\begin{equation}
{\rm Prob}(Y_2|Y_1, F_1, F_2)
\end{equation}
is (1) well defined and, if so, (2) what this probability is equal to, for all $(Y_1,F_1)$ and $(Y_2,F_2)$, for all
pairs of regions $R_1$ and $R_2$.   We will now develop a framework which can do this.

\section{Three levels of physical compression}

\subsection{Preliminaries}

Consider the probability
\begin{equation}
{\rm Prob}(Y|F)
\end{equation}
This is the probability that we see some stack $Y$ given procedure $F$.   It is unlikely that this probability is
well defined since it is conditioned only on choices of knob settings and not on any actual outcomes.   Thus,
instead, we consider the probabilities
\begin{equation}
{\rm Prob}(Y_R|F_R, C_{V-R})
\end{equation}
where $R$ is a large region (one containing a substantial fraction of the cards in $V$), $Y_R$ and $F_R$ are the
outcome set and procedure, respectively, in $R$.  And $C_{V-R}$ is some condition on $Y\cap(V-R)$ and $F\cap(V-R)$
(i.e. some condition on what is seen and what is done in region $V-R$).   We will assume that the probabilities
${\rm Prob}(Y_R|F_R, C_{V-R})$ are well defined for all $Y_R$ and $F_R$.  We will restrict our attention to the
case where condition $C_{V-R}$ is true and then we will only consider what happens in region $R$.  We might think
of $C_{V-R}$ as corresponding to the conditions that go into setting up and maintaining a laboratory (for example,
setting up the lasers, ensuring that the blinds are kept down, etc.).  Since we are always taking $C_{V-R}$ to be
true we will drop it from our notation writing ${\rm Prob}(Y_R|F_R)$. This way of setting up the framework is not
ideally suited to the cosmological context (where there is not any external condition like $C_{V-R}$). Ways round
this are discussed in \cite{Hardy1}.

\subsection{First level physical compression}

We will develop this framework by employing three levels of physical compression.  The first level of physical
compression pertains to a single region $R_1$ (inside $R$ of course).  We can write
\begin{equation}
{\rm Prob}(Y_R|F_R)={\rm Prob}(Y_{R_1}\cup Y_{R-R_1}| F_{R_1}\cup F_{R-R_1})
\end{equation}
We will think of $(Y_{R-R_1}, F_{R-R_1})$, which happens in $R-R_1$ as a {\it generalised preparation} for what
happens in region $R_1$ (we call it a generalised preparation since it is not, in general, restricted to the past
of $R_1$ - rather it pertains to the past, the future, and to elsewhere in so much as these words have meaning in
the absence of definite causal structure).   Further, we will think of each $(Y_{R_1}, F_{R_1})$ as corresponding
to some (measurement outcome, measurement choice) in region $R_1$ - we label them with $\alpha_1$. Thus, we have
$\alpha_1 \Leftrightarrow (Y^{\alpha_1}_{R_1}, F^{\alpha_1}_{R_1})$.  We can now write the above probability as
\begin{equation}
p_{\alpha_1}\equiv{\rm Prob}(Y^{\alpha_1}_{R_1}\cup Y_{R-R_1}| F^{\alpha_1}_{R_1}\cup F_{R-R_1})
\end{equation}
We will now define the {\it state in region} $R_1$ associated with a generalised preparation in $R-R_1$ to be that
thing represented by any mathematical object which can be used to calculate $p_{\alpha_1}$ for all $\alpha_1$.
Given this definition one mathematical object which clearly suffices to represent the state is
\begin{equation}
{\bf P}(R_1) = \left( \begin{array}{c} \vdots \\ p_{\alpha_1} \\ \vdots \end{array} \right)
\end{equation}
We can write
\begin{equation}\label{RdotP}
p_{\alpha_1} = {\bf R}_{\alpha_1}(R_1)\cdot {\bf P}(R_1)
\end{equation}
where ${\bf R}_{\alpha_1}(R_1)$ is a vector which has a 1 in position $\alpha_1$ and 0's everywhere else.  Now, in
general, a physical theory will correlate these probabilities.  This means that they will be related to each other.
Hence, we should be able to specify the state by giving a shorter list of probabilities (than in ${\bf P}$) from
which all the other probabilities can be calculated.  This provides some physical compression (compression due to
the physical theory itself). In fact we can choose to stick with linear physical compression.  Thus, we write the
state as a just sufficient set of probabilities
\begin{equation}
{\bf p}(R_1) = \left( \begin{array}{c} \vdots \\ p_{l_1} \\ \vdots \end{array} \right)    ~~~~~~ l_1\in \Omega_1
\end{equation}
where there exist vectors ${\bf r}_{\alpha_1}(R_1)$ such that a general probability is given by the linear equation
\begin{equation}
p_{\alpha_1} = {\bf r}_{\alpha_1}(R_1)\cdot {\bf p}(R_1)
\end{equation}
Clearly this is possible since, as a last resort, we have (\ref{RdotP}). Since the probabilities in ${\bf p}(R_1)$
are just sufficient for this purpose, there must exist a set of $|\Omega_1|$ linearly independent states chosen
from the allowed set of states. We will call $\Omega_1$ the fiducial set in region $R_1$.  The choice of fiducial
set for a region is unlikely to be unique. This does not matter. We can choose one set and stick with it.  We have
just employed linear physical compression here.  It is possible that if we employed more general mathematical
physical compression (allowing non-linear functions) we could do better.  This does not really matter since we are
free to choose linear physical compression as the preferred form of physical compression.  In fact, it can easily
be proven that if we are able to form mixtures of states (as we can in quantum theory) then we cannot do better
than linear physical compression (this is not surprising since probabilities combine in a linear way when we form
mixtures).  It is worth noting that in first level physical compression we implement the label change
\begin{equation}
\alpha_1 \longrightarrow l_1
\end{equation}
as we go from the set of all $\alpha_1$'s to the fiducial set $\Omega_1$.  The exact form of the first level
physical compression is encoded in the vectors ${\bf r}_{\alpha_1}$ (since if we know these vectors we can undo the
physical compression).  We define the matrix
\begin{equation}
\Lambda_{\alpha_1}^{l_1} \equiv r^{\alpha_1}_{l_1}
\end{equation}
where $r^{\alpha_1}_{l_1}$ are the components of ${\bf r}_{\alpha_1}$.  The matrix $\Lambda_{\alpha_1}^{l_1}$ tells
us how to undo the first level physical compression.  This matrix is likely to be very rectangular (rather than
square).

\subsection{Second level physical compression}

Now we come to second level physical compression.  This applies to two or more disjoint regions and corresponds to
the physical compression that happens over and above the first level compression for the composite regions.
Consider just two disjoint regions for the moment, $R_1$ and $R_2$.
\begin{equation} p_{\alpha_1\alpha_2} = {\rm Prob}(Y^{\alpha_1}_{R_1}\cup
Y^{\alpha_2}_{R_2} \cup Y_{R-R_1-R_2}|F^{\alpha_1}_{R_1}\cup F^{\alpha_2}_{R_2} \cup F_{R-R_1-R_2})
\end{equation}
where $\alpha_1$ and $\alpha_2$ label measurement plus outcomes in regions $R_1$ and $R_2$ respectively.  Now we
can reason as before. The state for region $R_1\cup R_2$ is given by any mathematical object which can be used to
calculate all $p_{\alpha_1\alpha_2}$.  Employing first level linear physical compression as before we can write the
state as
\begin{equation}
{\bf p}(R_1\cup R_2) = \left( \begin{array}{c} \vdots \\ p_{k_1k_2}\\ \vdots \end{array} \right)
~~~~~~k_1k_2\in\Omega_{12}
\end{equation}
where
\begin{equation}\label{raadotpaa}
p_{\alpha_1\alpha_2} = {\bf r}_{\alpha_1\alpha_2}(R_1\cup R_2)\cdot {\bf p}(R_1\cup R_2)
\end{equation}
We will now prove that there always exists a choice of fiducial set $\Omega_{12}$ such that
\begin{equation}\label{centralresult}
\Omega_{12} \subseteq \Omega_1\times\Omega_2
\end{equation}
where $\times$ represents the cartesian product (e.g. $\{1,2\}\times \{5,6\}=\{ 15, 16, 25, 26\}$). This result is
central to the method employed in this paper. Second level physical compression is nontrivial when $\Omega_{12}$ is
a proper subset of $\Omega_1\times\Omega_2$. To prove (\ref{centralresult}) note that we can write
$p_{\alpha_1\alpha_2}$ as
\begin{eqnarray}
\lefteqn{ {\rm prob}(Y^{\alpha_1}_{R_1}\cup Y^{\alpha_2}_{R_2}\cup Y_{R-R_1-R_2}   |  F^{\alpha_1}_{R_1}\cup
F^{\alpha_2}_{R_2}\cup F_{R-R_1-R_2}) }
\qquad\qquad\qquad\qquad\qquad\qquad \nonumber\\
&=&{\bf r}_{\alpha_1}(R_1)\cdot {\bf p}_{\alpha_2}(R_1)  \nonumber \\
&=&\sum_{l_1\in\Omega_1} r^{\alpha_1}_{l_1}(R_1) p^{\alpha_2}_{l_1}(R_1)   \nonumber \\
&=&\sum_{l_1\in\Omega_1} r^{\alpha_1}_{l_1}(R_1) {\bf r}_{\alpha_2}(R_2)\cdot {\bf p}_{l_1}(R_2) \nonumber \\
&=& \sum_{l_1l_2\in\Omega_1\times\Omega_2} r^{\alpha_1}_{l_1} r^{\alpha_2}_{l_2} p_{l_1l_2} \label{mainproof}
\end{eqnarray}
where ${\bf p}_{\alpha_2}(R_1)$ is the state in $R_1$ given the generalised preparation $(Y^{\alpha_2}_{R_2}\cup
Y_{R-R_1-R_2}, F^{\alpha_2}_{R_2}\cup F_{R-R_1-R_2})$ in region $R-R_1$, and ${\bf p}_{l_1}(R_2)$ is the state in
$R_2$ given the generalised preparation $(Y^{l_1}_{R_1}\cup Y_{R-R_1-R_2}, F^{l_1}_{R_1}\cup F_{R-R_1-R_2})$ in
region $R-R_2$ and where
\begin{equation}
p_{l_1l_2} = {\rm prob}(Y^{l_1}_{R_1}\cup Y^{l_2}_{R_2} \cup Y_{R-R_1-R_2} | F^{l_1}_{R_1}\cup F^{l_2}_{R_2} \cup
F_{R-R_1-R_2})
\end{equation}
Now we note from (\ref{mainproof}) that $p_{\alpha_1\alpha_2}$ is given by a linear sum over the probabilities
$p_{l_1l_2}$ where $l_1l_2\in\Omega_1\times\Omega_2$.   It may even be the case that we do not need all of these
probabilities.  Hence, it follows that $\Omega_{12}\subseteq\Omega_1\times \Omega_2$ as required.

We will now explain second level physical compression.  This is the physical compression that happens for a
composite regions over and above first level physical compression for the component regions.   From
(\ref{raadotpaa},\ref{mainproof}) we have
\begin{eqnarray}
p_{\alpha_1\alpha_2} &=& {\bf r}_{\alpha_1\alpha_2}(R_1\cup R_2)\cdot {\bf p}(R_1\cup R_2) \nonumber \\
                     &=& \sum_{l_1l_2} r^{\alpha_1}_{l_1} r^{\alpha_2}_{l_2} p_{l_1l_2} \nonumber \\
                     &=& \sum_{l_1l_2} r^{\alpha_1}_{l_1} r^{\alpha_2}_{l_2} {\bf r}_{l_1l_2}\cdot {\bf
                              p}(R_1\cup R_2)                                            \nonumber
\end{eqnarray}
Since we can find a spanning set of linearly independent states ${\bf p}(R_1\cup R_2)$, we must have
\begin{equation}
{\bf r}_{\alpha_1\alpha_2}(R_1\cup R_2) = \sum_{l_1l_2} r^{\alpha_1}_{l_1} r^{\alpha_2}_{l_2} {\bf
r}_{l_1l_2}(R_1\cup R_2)
\end{equation}
We define
\begin{equation}
\Lambda_{l_1l_2}^{k_1k_2} \equiv r^{l_1l_2}_{k_1k_2}
\end{equation}
where $r^{l_1l_2}_{k_1k_2}$ is the $k_1k_2$ component of ${\bf r}_{l_1l_2}$.  Hence,
\begin{equation}\label{causcomps}
 r^{\alpha_1\alpha_2}_{k_1k_2} = \sum_{l_1l_2} r^{\alpha_1}_{l_1} r^{\alpha_2}_{l_2} \Lambda_{l_1l_2}^{k_1k_2}
\end{equation}
This equation tells us that if we know $\Lambda_{l_1l_2}^{k_1k_2}$ then we can calculate ${\bf
r}_{\alpha_1\alpha_2}(R_1\cup R_2)$ for the composite region $R_1\cup R_2$ from the corresponding vectors ${\bf
r}_{\alpha_1}(R_1)$ and ${\bf r}_{\alpha_2}(R_2)$ for the component regions $R_1$ and $R_2$.  Hence the matrix
$\Lambda_{l_1l_2}^{k_1k_2}$ encodes the second level physical compression (the physical compression over and above
the first level physical compression of the component regions).  We can use it to define the {\it causaloid
product}
\begin{equation}
{\bf r}_{\alpha_1\alpha_2}(R_1\cup R_2) = {\bf r}_{\alpha_1}(R_1)\otimes^{\Lambda} {\bf r}_{\alpha_2}(R_2)
\end{equation}
where the components are given by (\ref{causcomps}).   The causaloid product generalises and unifies the various
products for quantum theory discussed in Section \ref{explorationof} (though in the context of a more general
framework - we will show in Section \ref{formulatingquantum} how quantum theory fits into this framework).

We can implement second level physical compression for more than two regions by applying the same reasoning.  Thus,
for multi-region physical compression, we implement
\begin{equation}
l_1l_2\dots l_n \longrightarrow k_1k_2\dots k_n
\end{equation}
in going from $\Omega_1\times\Omega_2\times\dots\times\Omega_n$ to $\Omega_{12\dots n}$ where the matrix
\begin{equation}
\Lambda_{l_1l_2\dots l_n}^{k_1k_2\dots k_n}
\end{equation}
encodes the second level physical compression.

\subsection{Third level physical compression}

Finally, we come to third level physical compression.  We can consider all regions to be composite regions made
from elementary regions $R_x$, $R_{x'}$, $R_{x''}$, etc.  Then we generate the following set of $\Lambda$ matrices.
\begin{equation}
\left(
\begin{array}{ll}
\Lambda_{\alpha_x}^{l_x}                        & \text{for all}~ x \in {\cal O}_R  \\  \\
\lambda_{l_xl_{x'}}^{k_xk_{x'}}                 & \text{for all}~ x, x' \in{\cal O}_R \\  \\
\lambda_{l_xl_{x'}l_{x''}}^{k_xk_{x'}k_{x''}}   & \text{for all}~ x, x',x'' \in{\cal O}_R \\  \\
 ~~~\vdots                                      &~~~ \vdots
\end{array} \right)
\end{equation}
where ${\cal O}_R$ is the set of $x$ in region $R$.   Given these $\Lambda$ matrices we can calculate the ${\bf r}$
vectors for any measurement outcome for any region using the causaloid product.  Now, just as the probabilities are
related to one another by the physical theory (thus enabling first and second level physical compression), we might
expect that these $\Lambda$ matrices are related to one another enabling us to calculate all of them from a smaller
set. Hence, we expect to be able to enact a third level of physical compression where the object
\begin{equation}
{\bf \Lambda}\equiv ( \text{subset of }~\Lambda's | \text{RULES} )
\end{equation}
enables us to calculate an arbitrary lambda matrix from the given subset (where RULES are a set or rules for doing
this). Such third level physical compression is, indeed, possible.  In Sec.\ \ref{formulatingquantum} we will show
how it is enacted in quantum theory. We will call ${\bf \Lambda}$ the {\it causaloid} (because it contains
information about the propensities for different causal structures). This is the central mathematical object in
this paper. For any particular physical theory the causaloid is fixed (this is modulo certain qualifications
concerning what might be regarded as boundary conditions that come from the conditioning $C_{V-R}$, though these
issues will, most likely, go away once we are in a cosmological setting \cite{Hardy1}). In fact, once we know the
causaloid we can perform any calculation possible in the physical theory (see Sec.\ \ref{usingthe}). Consequently,
the causaloid can be regarded as a specification of a physical theory itself.

The third level physical compression is accomplished by using identities relating $\Lambda$ matrices.  We can use
these to calculate higher order $\Lambda$ matrices (having more indices and corresponding to larger regions) from
lower order ones when certain conditions on the $\Omega$ sets are satisfied.  We will state some identities of this
form without proof. First, when $\Omega$ sets multiply so do $\Lambda$ matrices.
\begin{equation}\label{identity1}
\Lambda_{l_x\cdots l_{x'}l_{x''}\cdots l_{x'''}}^{k_x\cdots k_{x'}k_{x''}\cdots k_{x'''}} = \Lambda_{l_x\cdots
l_{x'}}^{k_x\cdots k_{x'}}\Lambda_{l_{x''}\cdots l_{x'''}}^{k_{x''}\cdots k_{x'''}}~~~\text{if}~~~ \Omega_{x\cdots
x' x'' \cdots x'''} = \Omega_{x\cdots x'} \times \Omega_{x''\cdots x'''}
\end{equation}
Second, there exists a family of identities from which $\Lambda$ matrices for composite regions can be calculated
from some pairwise matrices (given certain conditions on the $\Omega$ sets).  The first of this family is
\begin{equation}\label{identity2}
\Lambda_{l_1l_2l_3}^{k_1k_2k_3}=\sum_{k'_2\in\Omega_{2\not{\,3}}}\Lambda_{l_1k'_2}^{k_1 k_2}
\Lambda_{l_2l_3}^{k'_2k_3} ~~~{\rm if}~~~ \Omega_{123}= \Omega_{12}\times \Omega_{\not{\,2} 3} ~~~{\rm and}~~~
\Omega_{23}=\Omega_{2\not{\,3}}\times\Omega_{\not{\,2} 3}
\end{equation}
where the notation $\Omega_{\not{\,2} 3}$ means that we form the set of all $k_3$ for which there exists
$k_2k_3\in\Omega_{23}$.  The second in this family of identities is
\begin{equation}\label{identity2b}
\Lambda_{l_1l_2l_3l_4}^{k_1k_2k_3k_4}=\sum_{k'_2\in\Omega_{2\not{\,3}}, k'_3\in\Omega_{3\not{\,4}}}
\Lambda_{l_1k'_2}^{k_1 k_2} \Lambda_{l_2k'_3}^{k'_2k_3} \Lambda_{l_3l_4}^{k'_3k_4} ~~~{\rm if}~~
\begin{array}{l}
\Omega_{1234}= \Omega_{12}\times \Omega_{\not{\,2} 3} \times \Omega_{\not{\,3} 4} \\
\Omega_{23}=\Omega_{2\not{\,3}}\times\Omega_{\not{\,2} 3}  \\
 \Omega_{34}=\Omega_{3\not{\,4}}\times\Omega_{\not{\,3} 4}
\end{array}
\end{equation}
and so on.  These identities are elementary to prove (see \cite{Hardy1}).

\subsection{Using the causaloid to calculate correlations}\label{usingthe}

Once we have the causaloid, we can use it to calculate any ${\bf r}_{\alpha_1}(R_1)$ for any $\alpha_1$ and for any
region (whether composite or elementary) by using the causaloid product (using $\Lambda_{\alpha_x}^{l_x}$ from
first level physical compression to get the components of the ${\bf r}_{\alpha_x}(R_x)$ vectors for the elementary
regions to get us started). The causaloid can be used to calculate conditional probabilities as we require of the
formalism. Note
\begin{equation}
p\equiv {\rm Prob}(Y_1^{\alpha_1}|Y_2^{\alpha_2}, F_1^{\alpha_1}, F_2^{\alpha_2})= \frac{{\bf
r}_{\alpha_1\alpha_2}(R_1\cup R_2) \cdot {\bf p}(R_1\cup R_2)}{ \sum_{\beta_2}{\bf r}_{\alpha_1\beta_2}(R_1\cup
R_2) \cdot {\bf p}(R_1\cup R_2)}
\end{equation}
where $\beta_2$ runs over all outcomes for the measurement associated with $\alpha_2$ (recall that $\alpha_2$
labels a particular outcome of a particular measurement).  Therefore
\begin{enumerate}
\item $p$ is well defined iff
\begin{equation}
{\bf r}_{\alpha_1\alpha_2}(R_1\cup R_2) ~~\text{is parallel to} ~~ \sum_{\beta_2}{\bf r}_{\alpha_1\beta_2}(R_1\cup
R_2)
\end{equation}
because this is the only way for the probability to be independent of the state ${\bf p}(R_1\cup R_2)$ (as it must
since the state is associated with a generalised preparation outside $R_1\cup R_2$) since there exists a linearly
independent spanning set of such states.
\item If $p$ is well defined then it is given by
\begin{equation}
{\bf r}_{\alpha_1\alpha_2}(R_1\cup R_2) = p\sum_{\beta_2}{\bf r}_{\alpha_1\beta_2}(R_1\cup R_2)
\end{equation}
(i.e. equal to the ratio of the lengths of the vectors).
\end{enumerate}
This works for any pair of regions.  Hence, if we know the causaloid we can calculate whether any probability is
well defined and we can calculate its value if it is - this is the task we set ourselves at the end of Sec.
\ref{collectionand}.

\section{Formulating quantum theory in the causaloid framework}\label{formulatingquantum}

We will show that the theory for an arbitrary number of pairwise interacting qubits can be formulated within this
framework. Universal quantum computation can be carried out with such a system and so we will regard this as being
general enough for our purposes.  First, consider a single quantum system (which may be a qubit) acted up on by a
sequence of transformations/measurements labelled by $t=1, 2, \dots, T$. We can visualise this as a sequence of
boxes where each box has a knob for setting, $F(t)$, of the particular measurement being implemented and some
meters which record the outcome $s_t$ of the measurement. We record $(t, F(t), s_t)$ on a card for each $t$.  In
quantum theory such a measurement/transformation is associated with a set of completely positive trace
non-increasing linear maps (or superoperators) $\{ \$_{(t, F(t), s_t)} \} $ such that $\sum_{s_t} \$_{(t, F(t),
s_t)}$ (the sum is over all outcomes associated with a given measurement choice and a given $t$) is trace
preserving.  In our previous notation, $t$ plays the role of $x$, the elementary regions are $R_t$ (equal to the
set of all cards that can have $t$ on them), and $(Y_t,F_t)$ corresponds to $(\text{outcome}, \text{ setting})$ in
$R_t$. Further, we label each possible $(Y_t, F_t)$ by $\alpha_t$ in accordance with our previous notation.
Superoperators act on the input state to produce an output state
\begin{equation}
\rho(t+1) = \$_{\alpha_t} (\rho(t))
\end{equation}
Two important examples of superoperators are the unitary map $\rho \rightarrow U\rho U^\dagger $ (which preserves
the trace) and the projection map $ \rho \rightarrow\hat{P}\rho \widehat{P}$ (which decreases the trace in
general). In general, the probability of seeing the sequence of outcomes $s_1, s_2, \dots s_T$, given some
procedure $F(t)$, is given by
\begin{multline}\label{proballqt}
{\rm prob}(Y_T, Y_{T-1},\dots Y_1|F_T, F_{T-1},\dots F_1, \rho(0))  \\ = {\rm
trace}[\$_{\alpha_T}\circ\$_{\alpha_{T-1}}\circ\dots\circ\$_{\alpha_1}(\rho(0))]
\end{multline}

Now let us consider one elementary region $R_t$. We will write the probability in (\ref{proballqt}) as
\begin{equation}\label{palphatqt}
p_{\alpha_t} = {\rm trace}[\$_{T}\circ\dots \circ\$_{\alpha_t}\circ\dots\circ\$_{1}(\rho(0))]
\end{equation}
where we have suppressed $\alpha$'s from our notation except at the crucial time $t$.  Now note that, since
superoperaters are linear, we can expand a general superoperator in terms of a linearly independent fiducial set.
We will label the fiducial set by $l_t\in\Omega_t$ (we have $|\Omega_t|=N^4$ where $N$ is the dimension of the
Hilbert space for the system under consideration). Thus, we can write
\begin{equation}\label{texpanqt}
\$_{\alpha_t} = \sum_{l_t} r^{\alpha_t}_{l_t} \$_{l_t}
\end{equation}
where $\$_{l_t}$ is the fiducial set (this is not a unique choice).  Putting this into (\ref{palphatqt}) gives
\begin{equation}
p_{\alpha_t} = {\bf r}_{\alpha_t}\cdot {\bf p}
\end{equation}
where we are using our previous notation.  The $\Lambda$ matrices for the elementary regions are then given by
$\Lambda_{\alpha_t}^{l_t}= r^{\alpha_t}_{l_t}$ obtained by solving the set of linear equations (\ref{texpanqt}).
This accomplishes first level physical compression  for the elementary regions $R_t$ for a single quantum system
going from label $\alpha_t$ to label $l_t$.

Now we will write the probability in (\ref{proballqt}) as
\begin{equation}\label{palphatprimetqt}
p_{\alpha_{t'}\alpha_t} = {\rm trace}[\$_{T}\circ\dots\circ\$_{\alpha_{t'}}\circ\dots
\circ\$_{\alpha_t}\circ\dots\circ\$_{1}(\rho(0))]
\end{equation}
where we have suppressed $\alpha$'s from our notation except at times $t$ and $t'>t$.  If $t'=t+1$ then these two
times are immediately sequential. For a reason that will soon become apparent, we will choose the first member of
each fiducial set of superoperators to be equal to the identity map so we have $\$_1=I$ (where $I$ is the identity
map). Then we can write
\begin{equation}\label{tprimetexpanqt}
\$_{\alpha_{t'}}\circ\$_{\alpha_t} = \sum_{l_t} r^{\alpha_{t'}\alpha_t}_{1k_t} I\circ\$_{k_t}
\end{equation}
since the composition of two superoperators using $\circ$ is a map on $\rho$ and lives in the same space as a
single superoperator and so we can expand the composition in terms of a fiducial set of linearly independent
superoperators at one time. This means that
\begin{equation}\label{qtmultisequence}
\Omega_{t't}= \{ 1\} \times\Omega_t ~~~~\text{if} ~~t'=t+1
\end{equation}
and we see that we have non-trivial physical compression.   The $\Lambda$ matrices for this second level physical
compression of pairs of sequential elementary regions are given by
\begin{equation}
\Lambda_{l_{t'}l_t}^{1k_t} = r^{l_{t'}l_t}_{1k_t}
\end{equation}
by solving (\ref{tprimetexpanqt}).   The same technique works when we have any number of immediately sequential
regions. For three immediately sequential regions we have
\begin{equation}\label{qtomegamultisequence}
\Omega_{t''t't}= \{ 1\} \times\{ 1\} \times\Omega_t  ~~~~\text{if} ~~t''=t'+1=t+2
\end{equation}
and so on.

In the case that we have non-sequential times $t$ and $t'$ there is no physical compression and
\begin{equation}
\Omega_{t't}=\Omega_{t'}\times\Omega_{t} ~~~~\text{if} ~~t'>t+1
\end{equation}
Proof of this requires careful consideration of the form of (\ref{palphatprimetqt}) above.  We will omit this proof
here.  However, the physical reason for this is that different choices of intervening superoperators break the
possibility of any tight correlations between the two regions and so there is no physical compression.   The same
is true for any two clumps of regions with a gap.
\begin{equation}\label{multiplingomegas}
\Omega_{t'''\dots t''t'\dots t}=\Omega_{t'''\dots t''}\times\Omega_{t'\dots t} ~~~~\text{if} ~~t''>t'+1
\end{equation}

We now come to third level physical compression.  We can implement third level physical compression by noticing the
following. First note that we can divide any composite region into a set of regions  which we will call \lq\lq
clumps" where the regions in each clump are immediately sequential, and where there are gaps between the clumps.
Now note that (\ref{qtmultisequence}) (and its generalisations, such as (\ref{qtomegamultisequence}) to any number
of immediately sequential regions) satisfies the conditions on $\Omega$ sets such that identity (\ref{identity2})
(and its generalisations such as (\ref{identity2b})) hold. Hence, for each clump of immediately sequential regions
we can calculate the $\Lambda$ matrix employing this family of identities using just the $\Lambda$ matrices for
pairs of immediately sequential regions. Secondly, we see that (\ref{multiplingomegas}) satisfies the condition for
identity (\ref{identity1}) to hold - so that we can simply multiply the $\Lambda$ matrices from each clump to get
the $\Lambda$ matrix we are looking for. We will call this method the \lq\lq clumping method". This means that we
can write the causaloid for a single system in quantum theory as
\begin{equation}
{\bf \Lambda}=( \Lambda_{\alpha_t}^{l_\tau}, \Lambda_{l_{\tau+1}l_\tau}^{k_{t+1}k_t}|\text{RULES}=\text{clumping
method})
\end{equation}
where $\tau$ is some particular time $t$ (we only need specify these matrices for one $\tau$ since they will be the
same for all other $t$ by symmetry).

\begin{figure*}[p!]
\resizebox{\textwidth}{!} {\includegraphics{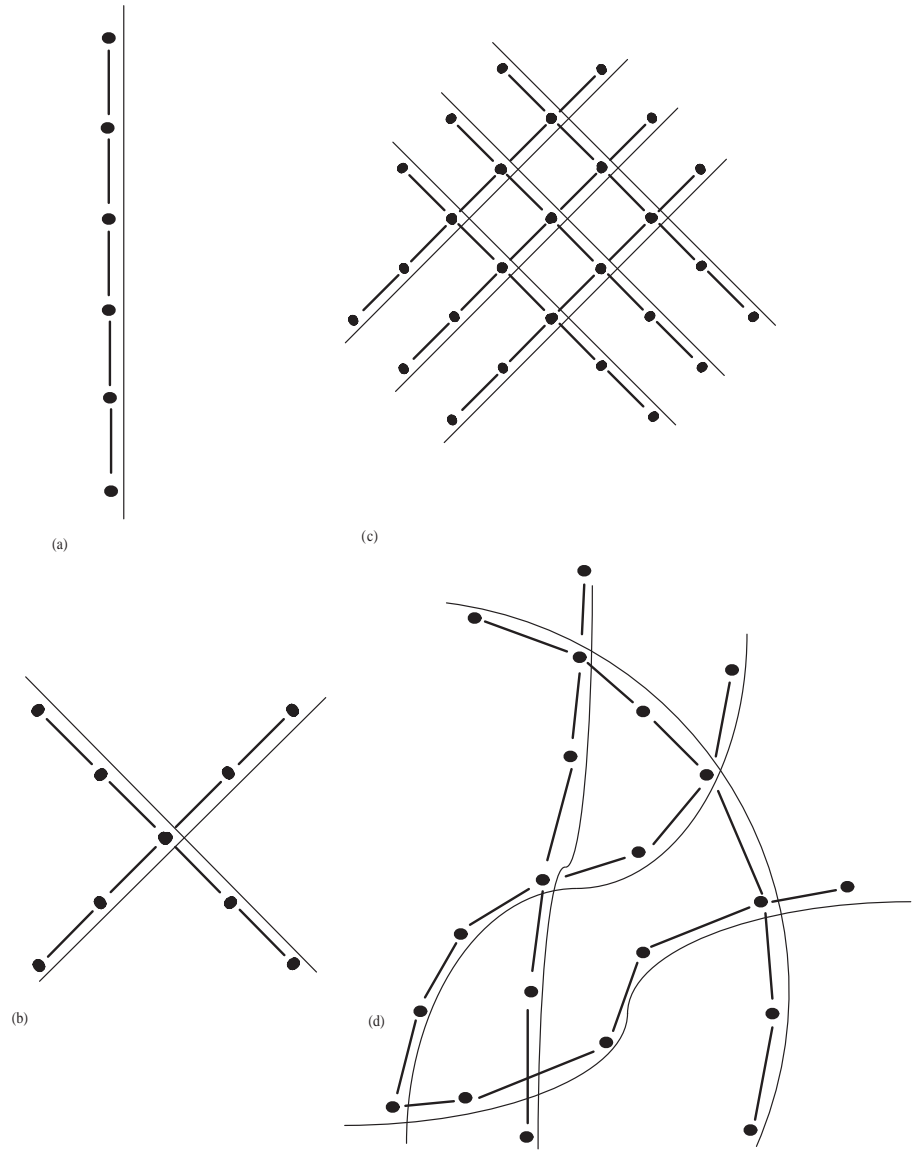}} \caption{\small Diagrams for (a) a single system (b) two
interacting systems, and (c,d) a number of systems interacting.}\label{cdiagram}
\end{figure*}

Now we will consider pairwise interacting qubits.  Examples of such pairs of interactions are given in Fig.\
\ref{cdiagram} (we will call these causaloid diagrams). Let each qubit be labelled by $i$.  The qubits are shown by
the thin lines. The nodes represent the elementary regions.  If two qubits pass through a node then they can
interact in that elementary region.  Nodes are labelled by $x$. Adjacent nodes (between which a qubit passes) are
represented by links. If we consider a single qubit $i$ then a sequence of times for this qubit is associated with
the sequence of labels $x$ along the thin line. We can build up the causaloid for this system of interacting qubits
by extending the methods above. To do this consider a node, $x$, at which two qubits, labelled by $i$ and $j$,
interact. We can act on these two qubits jointly with some measurement/transformation. This will be associated with
a set of superoperators $\$_{\alpha_x}$. A special subset of these superoperators are those that can be written in
tensor product form $\$^i_{\alpha_{xi}}\otimes\$^j_{\alpha_{xj}}$ where $\alpha_x\equiv(\alpha_{xi},\alpha_{xj})$
in these cases.  A subset of these are $\$^i_{l_{xi}}\otimes\$^j_{l_{xj}}$ where $l_{xi}\in\Omega_{xi}$ labels a
fiducial set of linearly independent superoperators on qubit $i$, and similarly for $j$.   Now, it turns out that
this particular set of product form superoperators form a complete linearly independent set for the general
superoperators on the two qubits.  That is, we can write
\begin{equation}\label{qtcoupling}
\$_{\alpha_x}=\sum_{l_{xi}l_{xj}\in\Omega_{xi}\times\Omega_{xj}}
r^{\alpha_x}_{l_{xi}l_{xj}}\$^i_{l_{xi}}\otimes\$^j_{l_{xj}}
\end{equation}
This means we can use fiducial measurements for which the qubits effectively decouple.  For each qubit we can apply
the clumping method to find the causaloid for that qubit.  Since the qubits effectively decouple for the fiducial
measurements, the $\Omega$ sets for composite regions involving more than one qubit will factorise between the
qubits.  Hence, a general $\Lambda$ matrix involving more than one qubit can be obtained by multiplying the
corresponding $\Lambda$ matrices for each qubit.  Then, to couple the qubits, we need only add the full
specification of the local lambda matrices
\begin{equation}
\Lambda_{\alpha_x}^{l_{xi}l_{xj}}=r^{\alpha_x}_{l_{xi}l_{xj}}
\end{equation}
which can be calculated from (\ref{qtcoupling}).   Hence, the causaloid is given by
\begin{equation}
{\bf \Lambda}=\left(\{\Lambda_{\alpha_x}^{l_{xi}l_{xj}} ~\forall~ x\},
\{\Lambda_{l_{xi}l_{x'i}}^{k_{xi}k_{x'i}}~\forall~ \text{adjacent} ~x,x' \} \left|  \begin{array}{l} \text{clumping
method}\\ \text{causaloid diagram}\end{array}\right. \right)
\end{equation}
 Note,
if a node only has one qubit passing through it then we list $\Lambda_{\alpha_x}^{l_{xi}}$ rather than
$\Lambda_{\alpha_x}^{l_{xi}l_{xj}}$. There is quite considerable physical compression at the third level.  If there
are $M$ nodes, then we only need list of order $M$ matrices (and these are low order matrices having only a small
number of indices) even though the number of possible $\Lambda$ matrices grows exponentially with $M$. We will,
most likely, be able to obtain further third level physical compression since symmetry considerations will mean
that we do not have to list separately the $\Lambda$ matrices for all $x$ and for all adjacent $x, x'$.

\section{Ideas on how to formulate General Relativity in the Causaloid framework}

General relativity has not yet been put into the causaloid framework. Such a formulation of GR would be
operational.  One idea is to pursue a line of thought suggested by Einstein.   He says
\begin{quote}
If, for example, events consisted merely in the motion of material points, then ultimately nothing would be
observable but the meetings of two or more of these points \cite{Einstein}.
\end{quote}
Thus, the data written onto a card would be a list of particles (assume each of these particles is labelled) which
are proximate.  We would collect many such cards forming a stack. The purpose of the physical theory would be to
correlate the data on these cards - and hence we would expect the causaloid formalism to work for this purpose.
There are a few problems with this approach.  Einstein introduces metric notions and it is not clear how this could
be recovered merely by looking at sets of coincidences.  One possible way to solve this problem would be to equip
each point particle with a clock and record the time of each particle's clock on the card also.  Another problem is
that the causaloid formalism is discrete rather than continuous.  There are discrete formulations of GR
\cite{Regge}, \cite{PullinGambini} but these tend to be in the canonical picture.  Nevertheless, we would probably
be satisfied with a discrete formulation of GR in the causaloid framework - especially if it turns out that QG is,
itself, a discrete theory since then GR would just be the continuous limit of a discrete theory. Unlike GR, the
causaloid framework has a notion of agency (there are knob settings).  However, no agency is a special case of
agency (where there is only one choice) so this need not be a problem.  Alternatively, we could try to recover the
notion of agency in GR.  For example, we could consider tiny differences in the matter distribution (such as those
in the brain) which are below the resolution of our experiment to be magnified so they are above the resolution.
This could be modelled in GR.

The theory we really want is what might be called probabilistic GR (ProbGR).  This would be to GR what statistical
mechanics is to Newtonian mechanics.  One problem with formulating ProbGR is that normally, when we formulate a
statistical version of a deterministic theory, we take a mixture of definite states accross space at a definite
time.  However, this would require a $3+1$ splitting against the spirit of GR and certainly against the spirit of
QG.  However, the causaloid framework would be a natural setting for ProbGR without introducing any such splitting.

\section{Ideas on how to formulate quantum gravity in the causaloid framework}

There are two strategies we might adopt to find a theory of QG in the causaloid framework.  First, we could
formulate both QT and ProbGR in this framework and then hope that some way of combining the essential features of
the two theories presents itself.   The \lq\lq map" that takes us from CProbT to QT could be applied to ProbGR to
get QG. This approach might work.  However, from a conceptual point of view it is not necessarily so clean.  We are
taking two less fundamental theories as part of the process by which we obtain a more fundamental one.  An
alternative approach would be to attempt to derive a theory of QG within the causaloid framework from scratch by
invoking some deep principles.  For example, we might attempt to formulate the equivalence principle in a
sufficiently general way that it applies to the causaloid framework.  This is clearly a much more difficult route
to get started on.  In practice, some combination of these two approaches is most likely to be successful.  It is
likely that, by having the two less fundamental theories formulated in the same framework, we will be in a better
position to extract principles from which QG can be derived.

\section{Conclusions}

A theory of QG is likely to have features that neither GR or QT have.  For example, in GR and QT there is a
definite matter of fact as to whether an interval is timelike or not (in QT this is specified in advance whereas in
GR we know this only after solving the equations).  The strategy we have adopted to work towards the construction
of QG is to construct a framework, the causaloid formalism, which is likely to be general enough to contain QG as a
special case. This is essential since if we work in a framework that cannot, in principle, contain QG then we have
no chance of formulating QG in the given framework.  The causaloid formalism does contain QT and it is likely to
contain GR.

The formulation of QT in this framework uses a notion of \lq\lq generalised preparation".  An example of this is
pre- and post-selection in the framework of Aharanov, Bergmann, and Lebowitz (ABL) \cite{ABL}.

In QG it is likely that we will lose the notion of an external time unaffected by what happens in the experiment.
This is likely to imply that we cannot have unitary evolution.  More accurately, it is likely to imply that the
theory which results when we take that limiting case of QG that approximates QT will not quite have unitary
evolution.  This might be consistent with collapse models (such as those of Ghirardi, Rimmini and Weber \cite{GRW},
and Pearle \cite{Pearle}). The possibility of a connection between gravity and non-unitary evolution does, of
course, have a long history (see in particular \cite{Diosi}, \cite{Penrose}, and for a different take see
\cite{Pullincollapse}). However, the situation might actually be more subtle. It is possible that, unlike in
collapse models, the theory will remain time-symmetric (in so much as such a notion makes sense in the absence of
fixed causal structure) just as the formulation of ABL is time symmetric. Collapse models employ the notion of an
evolving state at a fundamental level whilst such a notion is unlikely to be fundamental in QG.  But since the
measurement problem is a fundamental problem, we would like its solution to be implicit in the fundamental
formulation of QG rather than just in the limiting case of QT. This raises deep questions concerning whether
collapse is the right way to solve the measurement problem.

\vspace{6mm}

\noindent{\Large\bf Dedication}

\vspace{6mm}

It is a great honour to dedicate this paper to Giancarlo Ghirardi. One lesson implicit in his work on collapse
models, and particularly taken to heart here, is that we should think of modifying quantum theory in a hope to go
beyond our present theories.  Only then can we hope for experimental discrimination between theories.

\end{document}